\documentclass[superscriptaddress,twocolumn,10pt,nofootinbib,floatfix]{revtex4-1}

\usepackage{amssymb}
\usepackage{amsmath}
\usepackage{appendix}
\usepackage{times}
\usepackage{graphicx}
\usepackage{subeqnarray}
\usepackage{bbold}
\usepackage{enumerate}
\usepackage{color}
\usepackage{bm}
\usepackage[colorlinks=true, letterpaper=true, pdfstartview=FitV, linkcolor=blue, citecolor=blue, urlcolor=blue]{hyperref}

\usepackage[normalem]{ulem}

\UseRawInputEncoding 
\usepackage{makecell}  
\usepackage{booktabs}
\usepackage{float}

\setcitestyle{square}

\makeatletter
\renewcommand\@biblabel[1]{#1.}
\makeatother

\usepackage{braket}

\begin{document}

\title{Giant Magneto-Optical Sch\"{a}fer-Hubert Effect in Two-Dimensional van der Waals Antiferromagnets \textit{M}PS$_3$ (\textit{M}=Mn, Fe, Ni)}

\author{Ping Yang}
\affiliation{Key Laboratory of Advanced Optoelectronic Quantum Architecture and Measurement (MOE), School of Physics, Beijing Institute of Technology, Beijing 100081, China}
\affiliation{Beijing Key Lab of Nanophotonics and Ultrafine Optoelectronic Systems, School of Physics, Beijing Institute of Technology, Beijing 100081, China}

\author{Wanxiang Feng}
\email{wxfeng@bit.edu.cn}
\affiliation{Key Laboratory of Advanced Optoelectronic Quantum Architecture and Measurement (MOE), School of Physics, Beijing Institute of Technology, Beijing 100081, China}
\affiliation{Beijing Key Lab of Nanophotonics and Ultrafine Optoelectronic Systems, School of Physics, Beijing Institute of Technology, Beijing 100081, China}

\author{Gui-Bin Liu}
\affiliation{Key Laboratory of Advanced Optoelectronic Quantum Architecture and Measurement (MOE), School of Physics, Beijing Institute of Technology, Beijing 100081, China}
\affiliation{Beijing Key Lab of Nanophotonics and Ultrafine Optoelectronic Systems, School of Physics, Beijing Institute of Technology, Beijing 100081, China}

\author{Guang-Yu Guo}
\affiliation{Department of Physics and Center for Theoretical Physics, National Taiwan University, Taipei 10617, Taiwan}
\affiliation{Physics Division, National Center for Theoretical Sciences, Taipei 10617, Taiwan}

\author{Yugui Yao}
\affiliation{Key Laboratory of Advanced Optoelectronic Quantum Architecture and Measurement (MOE), School of Physics, Beijing Institute of Technology, Beijing 100081, China}
\affiliation{Beijing Key Lab of Nanophotonics and Ultrafine Optoelectronic Systems, School of Physics, Beijing Institute of Technology, Beijing 100081, China}

\date{\today}

\begin{abstract}
	
	{\bf The recent discovery of long-range magnetic order in atomically thin films has triggered particular interest in two-dimensional (2D) van der Waals (vdW) magnetic materials.  In this paper, we perform a systematic theoretical study of the magneto-optical Sch\"{a}fer-Hubert effect (MOSHE) in 2D vdW antiferromagnetic \textit{M}PS$_3$ (\textit{M} = Mn, Fe, Ni) with multifold intralayer and interlayer magnetic orders.  The formula for evaluating the MOSHE in 2D magnets is derived by considering the influence of a non-magnetic substrate.  The MOSHE of monolayer and bilayer \textit{M}PS$_3$ are considerably large ($>2^{\circ}$), originating from the strong anisotropy of in-plane optical conductivity.  The Sch\"{a}fer-Hubert rotation angles are surprisingly insensitive to the orientations of the N\'{e}el vector, while the Sch\"{a}fer-Hubert ellipticities are identified to be a good criterion to distinguish different interlayer magnetic orders.  Our work establishes a theoretical  framework for exploring novel 2D vdW magnets and facilitates the promising applications of the 2D \textit{M}PS$_3$ family in antiferromagnetic nanophotonic devices. }

\end{abstract}

\maketitle

\vspace{0.3cm}
\noindent{\bf INTRODUCTION}

\noindent Two-dimensional (2D) van der Waals (vdW) magnetic materials have attracted emerging attention since the discovery of intrinsically long-range ferromagnetic (FM) order in Cr$_2$Ge$_2$Te$_6$ and CrI$_3$ atomic layers~\cite{C-Gong2017,B-Huang2017}.  The highly tunable magnetism and other exciting physical properties by electric gating~\cite{B-Huang2018} and strain engineering~\cite{B-Yang2019,B-Huang2020} offer them a promising potential for applications in magnetic sensor, storage, and spintronics.  Magneto-optical spectroscopy is a powerful non-contact technique for investigating 2D magnetic materials.  For 2D ferromagnets, magneto-optical Kerr effect (MOKE) signals a solid evidence of long-range FM order even down to monolayer limit~\cite{B-Huang2017}. Furthermore, first-principles calculations of MOKE in thin films~\cite{Katayama1988,Suzuki1992} provide a complementary avenue to characterize 2D FM materials~\cite{W-Feng2016,X-Zhou2017,Y-Fang2018,Gudelli2019}.  For 2D antiferromagnets that have zero net magnetization, the MOKE as a first-order effect is vanishing, and therefore the commonly used magneto-optical techniques are based on second-order effects.~\cite{Saidl2017,Z-Zheng2018,Grigorev2021}  One option is to probe the difference in absorption or reflectivity for linearly polarized lights parallel and perpendicular to the N\'{e}el vector, which is known as magnetic linear dichroism (MLD).  Another option is to probe the polarization rotation upon transmission and reflection, which are called magneto-optical Voigt effect~\cite{Voigt1908} and magneto-optical Sch\"{a}fer-Hubert effect (MOSHE)~\cite{Schafer1990}, respectively.  Since the second-order magneto-optical effects in magnetic materials are usually very weak, the characterization of 2D AFM order has long been considered extremely challenging.

Transition metal thiophosphates \textit{M}PS$_3$ (\textit{M} = Mn, Fe, Ni) are a representative family of 2D vdW materials that host multifold intrinsically intralayer AFM orders~\cite{K-Kim2019b,Lee2016,Lancon2018}.  In a recent experiment, Zhang et al.~\cite{Q-Zhang2021} observed large MLD in FePS$_3$ with zigzag-AFM order.  The large magneto-optical signals enable the detection of 2D AFM domain orientations~\cite{Q-Zhang2021,ZL-Ni2022} and the study of ultrafast spin dynamics~\cite{XX-Zhang2021}.  Subsequently, the tuning of MLD in FePS$_3$ was realized by coupling with optical-cavity~\cite{H-Zhang2022}, and the MLD at specific wavelength can be even enhanced to a near-unity (100\%) value.  Such an optically anisotropic 2D magnetic material is desirable for achieving densely integrated polarization selective devices.  To date, most of the reports on large linear dichroism and its tuning for 2D materials have been limited to those with in-plane anisotropic crystal structures, such as black phosphorus~\cite{H-Yuan2015,Biswas2021} and GeSe~\cite{X-Wang2017}.  By contrast, anisotropic 2D magnetic materials are more promising for the fast field-effect control since the magnetic orders are sensitive to external stimuli, e.g., magnetic~\cite{X-Wang2021} and strain~\cite{Z-Ni2021} fields.  These recent advances call for an exploration of more excellent 2D AFM magneto-optical materials, however, theoretical studies on the second-order magneto-optical effects in thin films remain absent yet.

In this work, we systematically investigate a representative second-order magneto-optical effect, MOSHE, in 2D vdW AFM \textit{M}PS$_3$ using first-principles calculation together with magnetic group analysis.  A theoretical formula for evaluating the MOSHE in
2D magnetic materials placed on a non-magnetic substrate is derived for the first time. 
The MOSHE in FePS$_3$ and NiPS$_3$ with the zigzag-AFM order are close to or even exceed to the magnitudes of first-order magneto-optical effects in conventional ferromagnets, especially the Sch\"{a}fer-Hubert (SH) rotation angle in bilayer NiPS$_3$ records up to 2.4$^\circ$.  We also find that the MOSHE is insensitive to the magnetization direction, and the SH ellipticity can be used to identify interlayer magnetic structures.  Our work deepens the understanding of MOSHE in 2D antiferromagnets and facilitate further exploration of novel AFM magneto-optical devices.

\begin{figure*}
	\includegraphics[width=1.75\columnwidth]{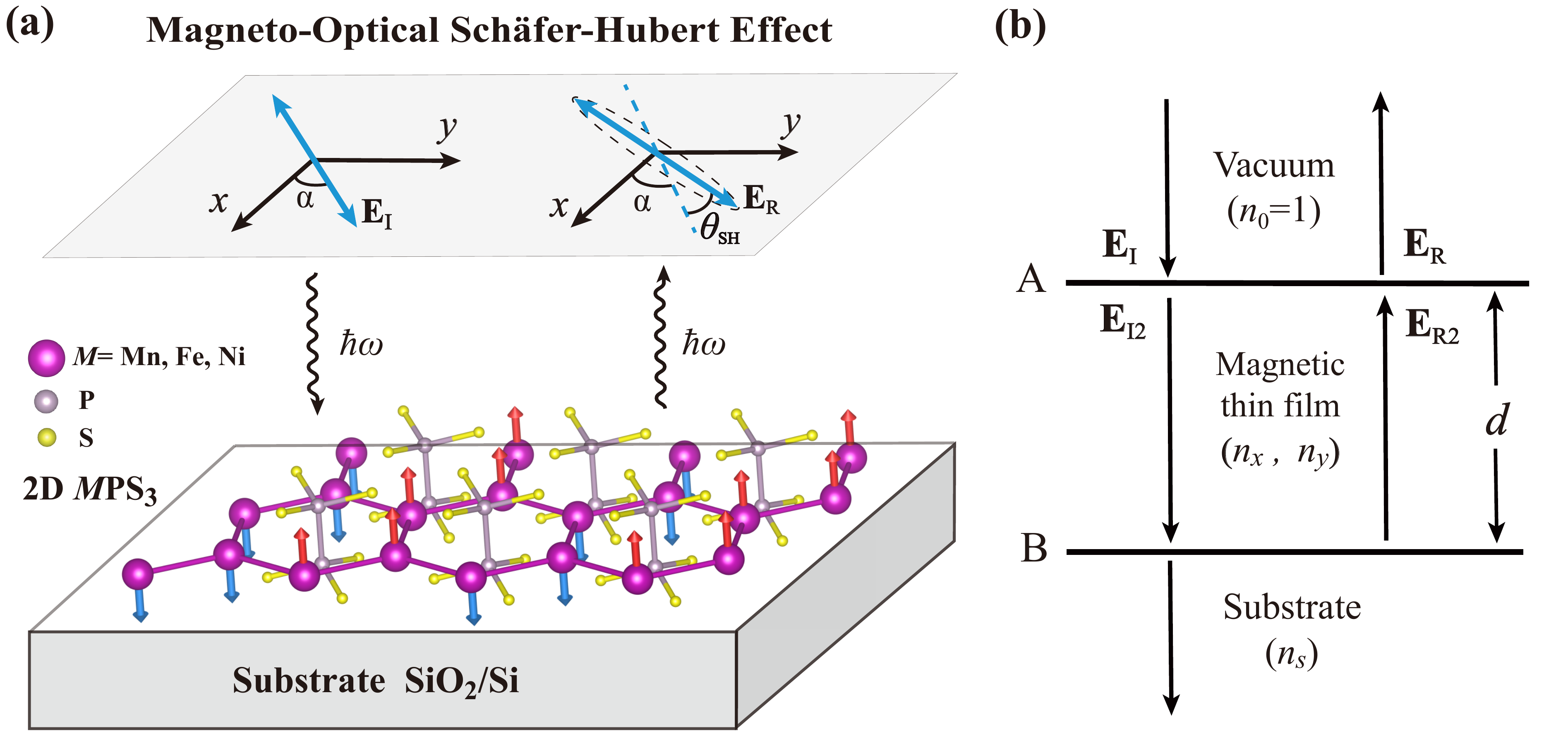}
	\caption{ (a) Schematic illustration of magneto-optical Sch\"{a}fer-Hubert effect emerged in 2D  antiferromagnets \textit{M}PS$_3$ (\textit{M} = Mn, Fe, Ni) prepared on SiO$_2$/Si substrate.  The incident light is linearly polarized with the electric field ($\textbf{E}_\text{I}$) orienting an angle of $\alpha$ from the optically anisotropic axis (here, $x$-axis). The reflected light becomes elliptically polarized and the polarization plane ($\textbf{E}_\text{R}$) deflects an angle of $\theta_{\rm SH}$ with respect to incident light ($\textbf{E}_\text{I}$).  (b) Optical paths in a magnetic thin film placed on an optically isotropic non-magnetic substrate.  Refractive indices ($n_0$, $n_x$, $n_y$, $n_s$) in each region and the electric fields ($\textbf{E}_\text{I}$, $\textbf{E}_\text{I2}$, $\textbf{E}_\text{R}$, $\textbf{E}_\text{R2}$) at the interface A are labeled, and $d$ denotes the thickness of magnetic thin film.}
	\label{fig:Optical path}
\end{figure*}

\vspace{0.3cm}
\noindent{\bf RESULTS AND DISCUSSION}

\noindent When a linearly polarized light normally shines (e.g., along the $z$-axis) on a thin film with in-plane magnetic anisotropy, the light propagating in the magnetic thin film can be decomposed into two polarized components along orthogonal anisotropic axes with different refractive indices ($n_{x}$, $n_{y}$) and reflectivity ($r_{x}$, $r_{y}$).  The reflected light would become elliptically polarized accompanied by a rotation of polarization plane with respect to the incident light, namely the MOSHE (Fig.~\ref{fig:Optical path}a).  If the electric field of incident light ($\textbf{E}_{\rm I}$) places at an angle of $\alpha = 45^{\circ}$ from the $x$-axis, the SH rotation angle ($\theta_{\rm SH}$) and ellipticity ($\psi_{\rm SH}$) reach up to their maximums, given by~\cite{Zvezdin1997}
\begin{equation}\label{SH-3D-1}
\begin{aligned}
\theta_{\rm SH}&=\frac{1}{2} \operatorname{atan}\left(\frac{2 \operatorname{Re} \chi}{1-|\chi|^{2}}\right) - \frac{\pi}{4}, \\ 
\psi_{\rm SH}&=\frac{1}{2} \operatorname{asin}\left(\frac{2 \operatorname{Im} \chi}{1+|\chi|^{2}}\right),
\end{aligned}	
\end{equation}	
where $\chi = r_{y}/r_{x}$.  The reflectivity of a magnetic thin film at the interface A (Fig.~\ref{fig:Optical path}b) can be written as
\begin{equation}\label{2D-reflect}
r_{x (y)}= \frac{n_{0}-\tilde{n}_{x (y)}}{n_{0}+ \tilde{n}_{x (y)}}.
\end{equation}
Here, $n_{0}=1$ is the refractive index of vacuum, $\tilde{n}_{x (y)}$ is the effective refractive index of a magnetic thin film by considering the influence of its substrate,
\begin{equation}\label{2D-refractive}
\tilde{n}_{x (y)}= \frac{1-r^{\prime}_{x (y)} \beta_{x (y)}}{1+r^{\prime}_{x (y)} \beta_{x (y)}}n_{x (y)},
\end{equation}
in which $\beta_{x (y)}= {\rm exp}(2i \omega d n_{x (y)}/c)$ with the light frequency $\omega $, light speed $c$, and film thickness $d$.  The reflectivity of substrate at the interface B is $r^{\prime}_{x(y)}=(n_{x (y)}-n_{s})/(n_{x (y)}+n_{s})$ and $n_{s}$ is the refractive index of substrate.  Plugging Eqs.~(\ref{2D-reflect}) and (\ref{2D-refractive}) into Eq.~(\ref{SH-3D-1}), the complex SH angle can be recast as
\begin{equation}\label{2D-SH-1}	
\theta_{\rm SH} + i \psi_{\rm SH} \approx \frac{\tilde{n}_{x}-\tilde{n}_{y}}{1-\tilde{n}_{x} \tilde{n}_{y}}.
\end{equation}

For monolayer and few-layer of 2D materials whose thicknesses are far less than the wavelength of visible light ($\lambda$), the effective refractive index can be approximated to $\tilde{n}_{x(y)} \approx n_{s}-i \cdot 2 \pi d (n_{x(y)}^{2}-n_{s}^{2})/ \lambda$.  In the case of conventional MOSHE induced by in-plane magnetization (e.g., along the $x$-axis), the refractive indices by solving the Fresnel equation are given by
$n_{x} =\sqrt{\epsilon_{xx}}$, $n_{y} =\sqrt{\epsilon_{yy}+\epsilon_{yz}^{2} / \epsilon_{zz}}$, in which $\epsilon_{\mu\nu}$ with $\mu,\nu\in\{x,y,z\}$ is the permittivity tensor.  Then, the complex SH angle can be simplified to
\begin{equation}\label{2D-SH}
	\theta_{\rm SH} + i \psi_{\rm SH} \approx  \frac{i\omega d} {c(n^2_s -1)}(\epsilon_{xx}-\epsilon_{yy}-\frac{\epsilon^2_{yz}}{\epsilon_{zz}}).
\end{equation}	 
We find that the complex SH angle can be related to the complex Voigt angle~\cite{Mertins2001} via
\begin{equation}\label{SH_vs_Voigt}
 \theta_{\rm SH} +i\psi_{\rm SH} = \frac{2(n_{x}+n_{y})}{1- n_s ^2} (\theta_{\rm V} -i\psi_{\rm V}),
\end{equation}
where $\theta_{\rm V}$ and $\psi_{\rm V}$ are Voigt ration angle and ellipticity, respectively.  If the substrate has a relatively small refractive index ($n_{s}\rightarrow1$), the SH angle will be much larger than the Voigt angle, indicating that the optical detection upon reflection is more suitable than upon transmission for studying the second-order magneto-optical effects of magnetic thin films.

\begin{table*}[t]
	\caption{Magnetic space groups of monolayer \textit{M}PS$_3$ with different magnetic orders.  The magnetization directions are labeled in brackets.  The symbol $\checkmark$ ($\times $) indicates the in-plane optically anisotropy (isotropy).  The dipole selection rules at some high-symmetry points (e.g., $\Gamma$ and K) are listed.}
	\label{tab:mag-group}
	\setlength{\tabcolsep}{7.0pt}
	\renewcommand{\arraystretch}{2.0}
	\begin{tabular}{*{4}{c}}
	\hline
	\hline
	Magnetic orders & \thead{Magnetic \\space group } & \thead{In-plane \\ anisotropy \\ ($\sigma_{xx} \neq \sigma_{yy} $)} & \thead{Dipole  \\ Sele. rules \\ ($\textbf{E} \perp z$)} \\
	\hline 
	FM ($x$)   & $ C2'/m' $  &  $\checkmark$  & \thead{$\Gamma_{2}^{+}\leftrightarrow \Gamma_{2}^{-}$ } \\
	FM ($z$)   & $ P\overline{3}1m' $  &  $\times $ & \thead{$\Gamma_{4}^+ \leftrightarrow \Gamma_{5}^- ,\Gamma_{6}^-$ \quad K$_{4} \leftrightarrow$  K$_{5}$ \\ $\Gamma_{5}^+ \leftrightarrow  \Gamma_{4}^- ,\Gamma_{6}^- $ \quad K$_{4} \leftrightarrow $ K$_{6}$\\ $\Gamma_{6}^+ \leftrightarrow  \Gamma_{4}^- ,\Gamma_{5}^-$ \quad K$_{5} \leftrightarrow $ K$_{6}$} \\
    N\'{e}el-AFM ($x$)  & $ C2'/m $   &  $\checkmark$ & \thead{$\Gamma_{3}\Gamma_{4}\leftrightarrow \Gamma_{3}\Gamma_{4}$}  \\
    N\'{e}el-AFM ($z$)   & $ P\overline{3}'1m $  &  $\times $ &\thead{$\Gamma_4 \leftrightarrow \Gamma_4 $\qquad $\mathrm{K}_{4} \leftrightarrow \mathrm{K}_{4}$  \\ $\Gamma_{4} \leftrightarrow \Gamma_{5}\Gamma_{6}$ \ \  $\mathrm{K}_4 \leftrightarrow \mathrm{K}_{5}\mathrm{K}_{6}$ } \\
	zigzag-AFM ($x$,$z$)   & $ P_{c}2_{1}/m $ &  $\checkmark$ & \thead{$\Gamma_{3}^{+}\Gamma_{4}^{+} \leftrightarrow \Gamma_{3}^{-}\Gamma_{4}^{-} $}  \\
	stripy-AFM ($x$,$z$)   & $ P_{a}2_{1}/c $ &  $\checkmark$  & \thead{$\Gamma_{3}^{+}\Gamma_{4}^{+} \leftrightarrow \Gamma_{3}^{-}\Gamma_{4}^{-}$} \\
	\hline
	\hline
\end{tabular}
\end{table*}

The complex SH angle (see Eq.~(\ref{2D-SH})) can also be written in terms of optical conductivity using the relationship between permittivity and optical conductivity, given by, $\epsilon_{\mu\nu}=\delta_{\mu\nu} + \frac{4\pi i}{\omega}\sigma_{\mu\nu}$.  The off-diagonal elements of the optical conductivity containing the $z$-component (e.g., $\sigma_{yz}$) have to be zero due to the 2D nature of our considered systems.  This can be read from Eq.~(\ref{eq:OPC}) since the quenched electron velocity along the $z$ direction ($\hat{\upsilon}_{z}=0$) leads to the vanishing $\sigma_{yz}$ and $\sigma_{zx}$.  Therefore, the complex SH angle is simply expressed as
\begin{equation}\label{2D-SH-optc}
	\theta_{\rm SH} + i \psi_{\rm SH} \approx \frac{4\pi d} {c(n^2_s-1)}(\sigma_{yy}-\sigma_{xx}),
\end{equation}
which is the formula implemented in our first-principles calculations.  The 2D vdW magnetic materials are often grown on transparent substrates, such as SiO$_2$, whose refractive index $n_s$ is a real number.  In this case, the SH rotation angle and ellipticity are determined by the real and imaginary parts of conductivity anisotropy (i.e., $\sigma_{yy}-\sigma_{xx}$), respectively.  On account of this relationship, the conductivity anisotropy can be accurately measured by MOSHE spectroscopy.

\begin{figure}[b]
	\includegraphics[width=1.0\columnwidth]{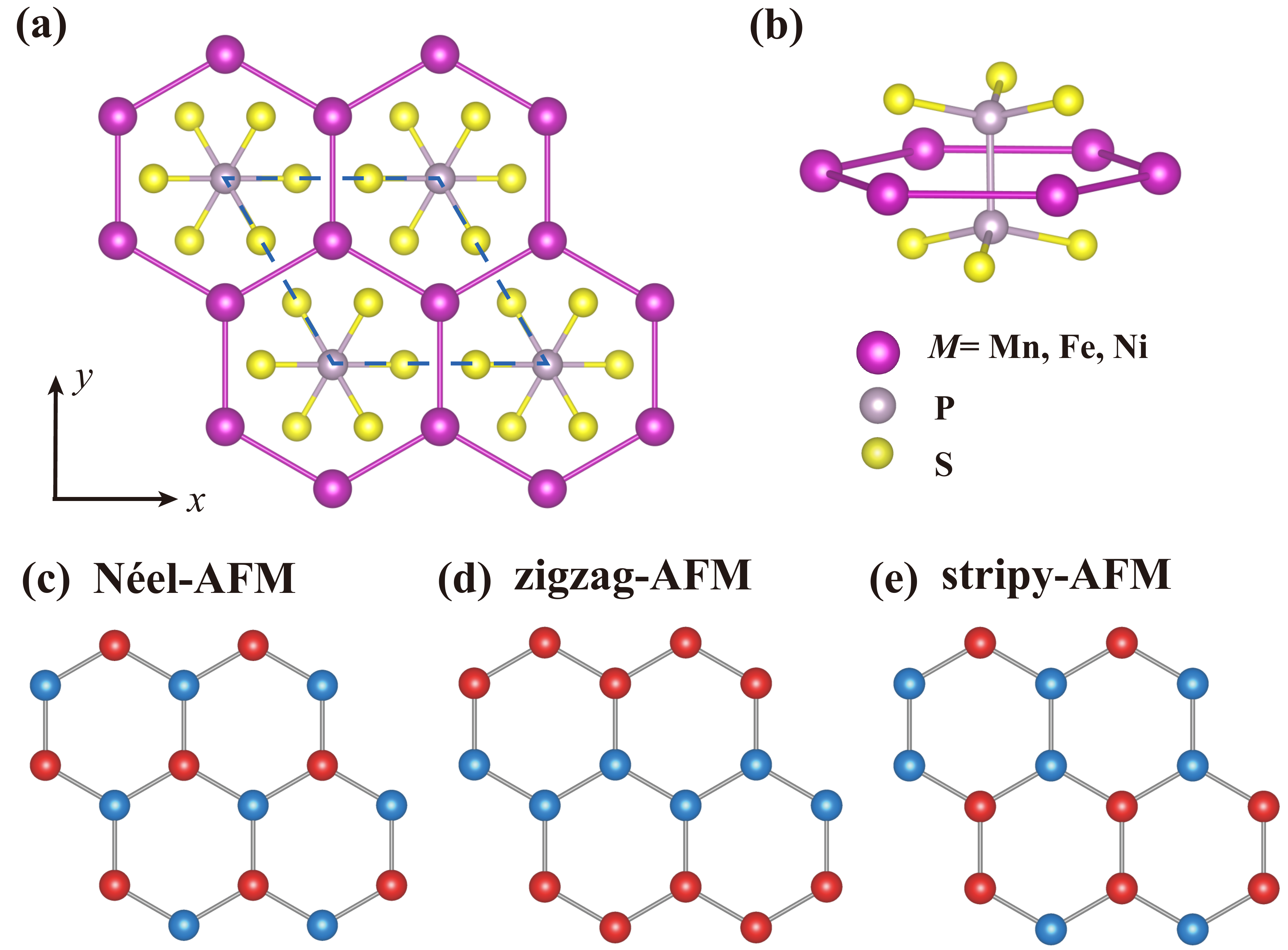}
	\caption{ (a,b) Top and side views of monolayer \textit{M}PS$_3$.  Blue dashed lines draw out the primitive cell of non-magnetic state.  (c-e) The N\'{e}el-, zigzag-, and stripy-antiferromagnetic orders on a honeycomb lattice.  Red and blue spheres represent the \textit{M} atoms with opposite directions of spin magnetic moments.}
	\label{fig:model}
\end{figure}

For monolayer \textit{M}PS$_{3}$, the transition metal atoms \textit{M} form a flat honeycomb lattice and a bipyramid of P$_2$S$_6$ ligand locates at the center of hexagon (Fig.~\ref{fig:model}a,b).  If removing the magnetic orders, monolayer \textit{M}PS$_{3}$ is in-plane isotropic due to its crystallographic point group of $D_{3d}$.  Nevertheless, the honeycomb lattice can host a variety of magnetic orders, including FM state as well as N\'{e}el-, zigzag-, and stripy-AFM states (Fig.~\ref{fig:model}c-e)~\cite{Sivadas2015}, depending on the relative strength of intralayer first, second, and third nearest-neighbour exchange interactions.  MnPS$_{3}$ displays the N\'{e}el-AFM order with the out-of-plane ($z$-axis) magnetic easy axis~\cite{K-Kim2019b}.  FePS$_{3}$ and NiPS$_{3}$ display the zigzag-AFM order with the out-of-plane ($z$-axis)~\cite{Lee2016} and in-plane ($x$-axis)~\cite{K-Kim2019a} magnetization, respectively.  The exfoliated atomic layers persist long-range AFM orders down to bilayer or even monolayer limit, and their magnetic critical temperatures are nearly independent on thickness.  Moreover, the magnetization for N\'{e}el- and zigzag-AFM states can be tuned between the out-of-plane and in-plane directions via atomic substitution~\cite{Basnet2022}, and the FM state was predicted to be their ground states under sufficient large carrier density~\cite{Chittari2016}.

\begin{figure*}[t]
	\centering
	\includegraphics[width=2.0\columnwidth]{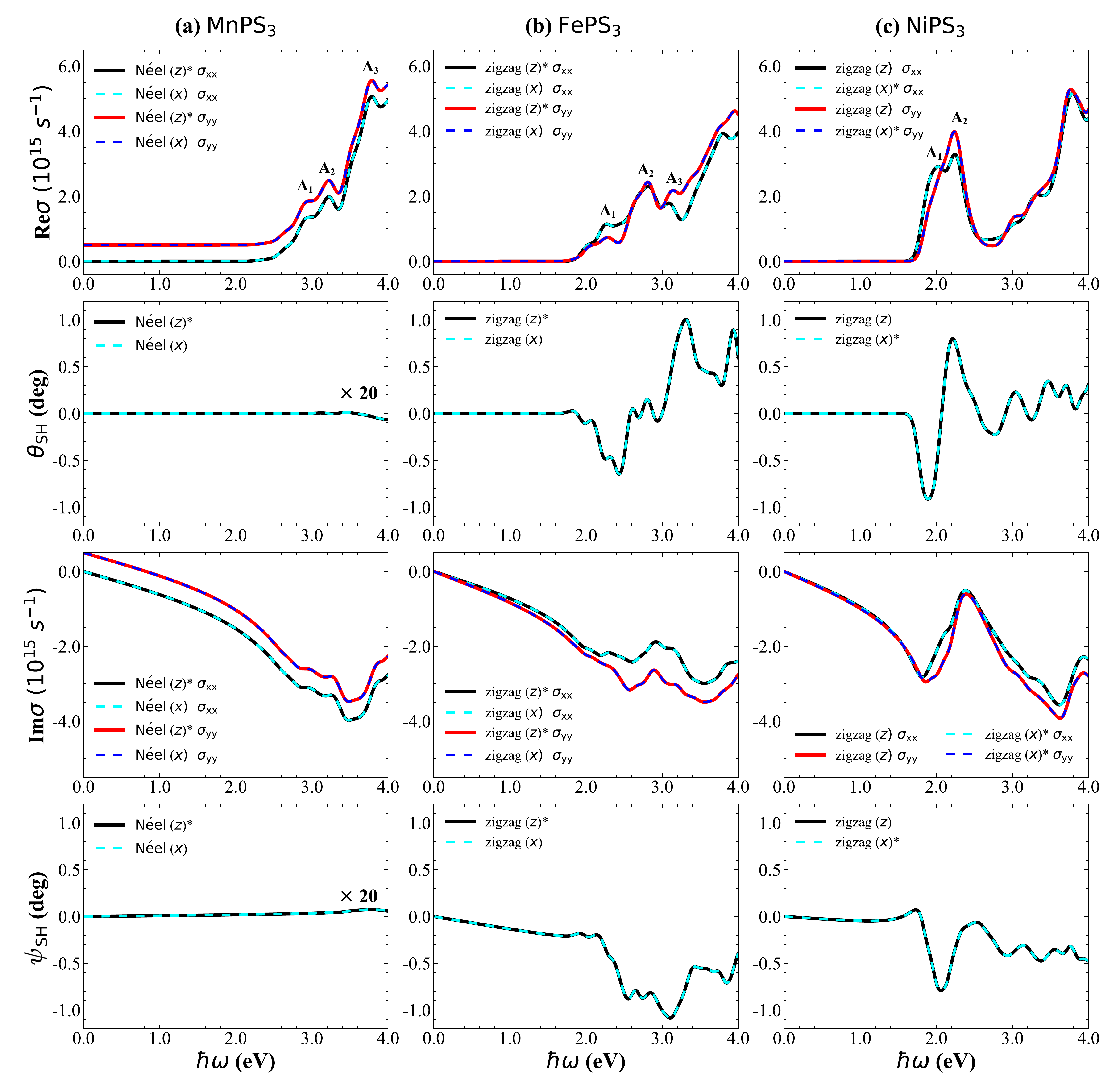}
	\caption{ Optical conductivities and MOSHE spectra of monolayer (a) MnPS$_3$, (b) FePS$_3$, and (c) NiPS$_3$ on SiO$_{2}$ substrate.  The panels from top to bottom show the real part of optical conductivity (Re$\sigma$), SH rotation angle ($\theta_{\rm SH}$), imaginary part of optical conductivity (Im$\sigma$), and SH ellipticity  ($\psi_{\rm SH}$), respectively.  The magnetization direction of each magnetic order is indicated in brackets, and an asterisk labels the ground state.  The Re$\sigma_{yy}$ and Im$\sigma_{yy}$ of MnPS$_3$ are moved upward by 0.5$\times$10$^{15}$s$^{-1}$ for a clear observation, and the $\theta_{\rm SH}$ and $\psi_{\rm SH}$ of MnPS$_3$ are multiplied by a factor of 20.  A$_1$, A$_2$, and A$_3$ mark several absorption peaks of Re$\sigma$ in the low-energy range.}
	\label{fig:SH_1L}
\end{figure*}

Before practically calculating MOSHE, we conduct symmetry analysis to evaluate which magnetic order breaks the in-plane optically isotropy of monolayer \textit{M}PS$_{3}$.  The magnetic space groups computed by \textsc{isotropy} code~\cite{isotropy} are listed in Table~\ref{tab:mag-group}, in which the shapes of optical conductivity tensors are identified by \textsc{symmetr} code~\cite{Zelezny2017a,Zelezny2018a}.  As expected, all of the magnetic orders with the magnetization along the $x$-axis are in-plane anisotropic, which allows the MOSHE.  For FM and N\'{e}el-AFM orders with the spins along the $z$-axis, the in-plane isotropy is preserved by the three-fold rotational symmetry in magnetic space groups of $P\overline{3}1m'$ and $P\overline{3}'1m$, respectively.  The magnetic space groups of zigzag- and stripy-AFM orders with the magnetization along the $z$-axis are the same as that along the $x$-axis, such that the $z$-axis magnetization is also in-plane anisotropic and may also lead to the MOSHE.  According to the mirror symmetry $\mathcal{M}_{y}$ in the zigzag-AFM order, the orthogonal anisotropic axes are determined to be the $x$- and $y$-axes as shown in Fig.~\ref{fig:model}.

\begin{figure*}[t]
	\centering
	\includegraphics[width=2.0\columnwidth]{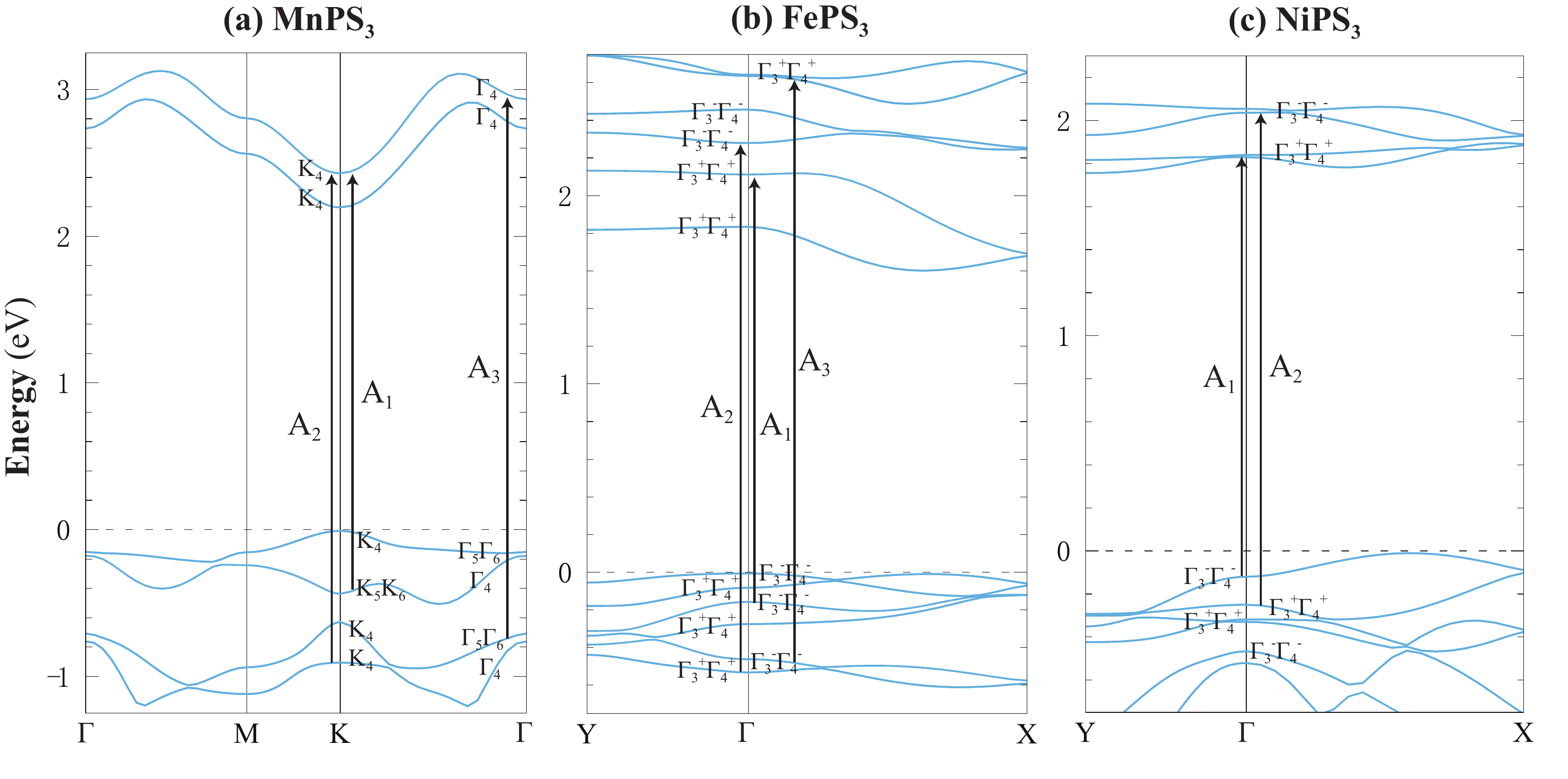}
	\caption{ Relativistic band structures of monolayer (a) MnPS$_3$ with the $z$-axis N\'{e}el-AFM order, (b) FePS$_3$ with the $z$-axis zigzag-AFM order, and (c) NiPS$_3$ with the $x$-axis zigzag-AFM order.  The irreducible representations of relevant bands at the $\Gamma$ and K points are labeled.  The principal interband transitions A$_1$, A$_2$, and A$_3$ are indicated by arrows, corresponding to the peaks of Re$\sigma_{xx}$ and Re$\sigma_{yy}$ in Fig.~\ref{fig:SH_1L}.}
	\label{fig:band}
\end{figure*}

Figure~\ref{fig:SH_1L} plots the calculated optical conductivities and MOSHE spectra of monolayer AFM \textit{M}PS$_{3}$.  We first discuss the results of each material on its magnetic ground state.  For MnPS$_{3}$ with the $z$-axis N\'{e}el-AFM order (Fig.~\ref{fig:SH_1L}a), the spectrum of $\sigma_{xx}$ is identical to that of $\sigma_{yy}$, which is governed by the in-plane optical isotropy, and the resulting SH rotation angle ($\theta_{\rm SH}$) and SH ellipticity ($\psi_{\rm SH}$) are negligibly small.  The absorptive parts of optical conductivity tensor, Re$\sigma_{xx}$ and Re$\sigma_{yy}$, are determined by the symmetry-allowed dipole selection rules listed in Table~\ref{tab:mag-group}, from which one can analyze the origination of main peaks in conductivity spectra.  For example, the A$_1$ and A$_2$ peaks at the energies of 2.9 eV and 3.2 eV originate from the interband transitions ${\rm K}_{5}{\rm K}_{6}\rightarrow {\rm K}_{4}$ and ${\rm K}_{4}\rightarrow {\rm K}_{4}$ at the K-point, respectively, and the A$_3$ peak at the energy of 3.7 eV originates from the interband transition $\Gamma_{5}\Gamma_{6}\rightarrow \Gamma_{4}$ at the $\Gamma$-point, as depicted in Fig.~\ref{fig:band}a.  For FePS$_{3}$ with the $z$-axis zigzag-AFM order (Fig.~\ref{fig:SH_1L}b), one can discern a clear anisotropy in real and imaginary parts of optical conductivity above the absorption edge ($\sim$1.8 eV).  The spectra of Re$\sigma_{xx}$ and Re$\sigma_{yy}$ feature three peaks of A$_1$, A$_2$, and A$_3$ at the energies of 2.3 eV, 2.8 eV, and 3.1 eV, respectively, which come from the interband transitions between the $\Gamma_{3}^{+}\Gamma_{4}^{+}$ and $\Gamma_{3}^{-}\Gamma_{4}^{-}$ states at the $\Gamma$-point (Fig.~\ref{fig:band}b).  The obvious difference in values between Re$\sigma_{xx}$ and Re$\sigma_{yy}$ around the A$_1$ and A$_3$ peaks generate the maximal SH rotation angles of -0.7$^{\circ}$ at 2.4 eV and of 1.0$^{\circ}$ at 3.3 eV, respectively.  The SH ellipticity is always negative and reaches up to -1.1$^{\circ}$ at 3.1 eV.  For NiPS$_{3}$ with the $x$-axis zigzag-AFM order (Fig.~\ref{fig:SH_1L}c), the real part of optical conductivity resembles the experimental detection of its bulk crystal~\cite{S-Kim2018}.  Both Re$\sigma_{xx}$ and Re$\sigma_{yy}$ spectra show the A$_2$ peak at 2.3 eV due to the interband transition $\Gamma_{3}^{+}\Gamma_{4}^{+}\rightarrow \Gamma_{3}^{-}\Gamma_{4}^{-}$, while an additional peak A$_1$ appears at 2.0 eV for Re$\sigma_{xx}$ which is related to the transition from the $\Gamma_{3}^{-}\Gamma_{4}^{-}$ state (highest valance band) to the $\Gamma_{3}^{+}\Gamma_{4}^{+}$ state (lowest conduction band) at the $\Gamma$-point (Fig.~\ref{fig:band}c).  In the energy range of 1.7 $\sim$ 2.5 eV, the significant anisotropy of optical conductivity has to result in large SH rotation angles, e.g., -0.9$^{\circ}$ at 1.9 eV and 0.8$^{\circ}$ at 2.2 eV.  The corresponding SH ellipticity is also obviously large with a peak of -0.8$^{\circ}$ at 2.1 eV.

Of particular interest here is that the optical conductivity spectra are almost not changed when the magnetization direction changes from the $z$-axis to the $x$-axis or vice versa (Fig.~\ref{fig:SH_1L}).  This is very similar to the cases of three-dimensional noncollinear AFM Mn$_3$\textit{X} (\textit{X} = Rh, Ir, Pt)~\cite{W-Feng2015} and 2D vdW FM Fe$_n$GeTe$_2$ ($n$ = 3, 4, 5)~\cite{X-Yang2021}.  It can be easily understood as the longitudinal optical conductivities ($\sigma_{xx}$ and $\sigma_{yy}$) are closely related to the joint density of states and interband transition probability~\cite{Stroppa2008} which are basically not influenced when the angle between adjacent spins keeps fixed.  It follows that the SH spectra of \textit{M}PS$_{3}$ are insensitive to magnetization direction, e.g., FePS$_{3}$ and NiPS$_{3}$ with the $z$- and $x$-axes zigzag-AFM orders (Fig.~\ref{fig:SH_1L}b,c).  In the case of MnPS$_{3}$ with the $x$-axis N\'{e}el-AFM order, the optical conductivities are identical to the results of $z$-axis N\'{e}el-AFM order, such that the SH rotation angel and ellipticity are also negligibly small (Fig.~\ref{fig:SH_1L}a), even though the appearance of MOSHE with the $x$-axis N\'{e}el-AFM order is allowed by symmetry (Table~\ref{tab:mag-group}).  Similarly, since the $z$-axis FM order exhibits in-plane isotropy, the MOSHE in all the three materials with the $x$-axis FM order are rather small (see Supplementary Fig.~1), e.g., the largest SH rotation angle (appearing in FePS$_{3}$) is only 0.05$^{\circ}$.  Therefore, we suggest that it is more likely to observe large second-order magneto-optical effects in AFM materials that exhibit in-plane anisotropy when the spins are out-of-plane oriented, such as the \textit{M}PS$_3$ family with the zigzag-AFM and stripy-AFM (Fig.~\ref{fig:SH_1L}b,c) orders (see Supplementary Fig.~2).

\begin{figure}[h]
	\centering
	\includegraphics[width=1.0\columnwidth]{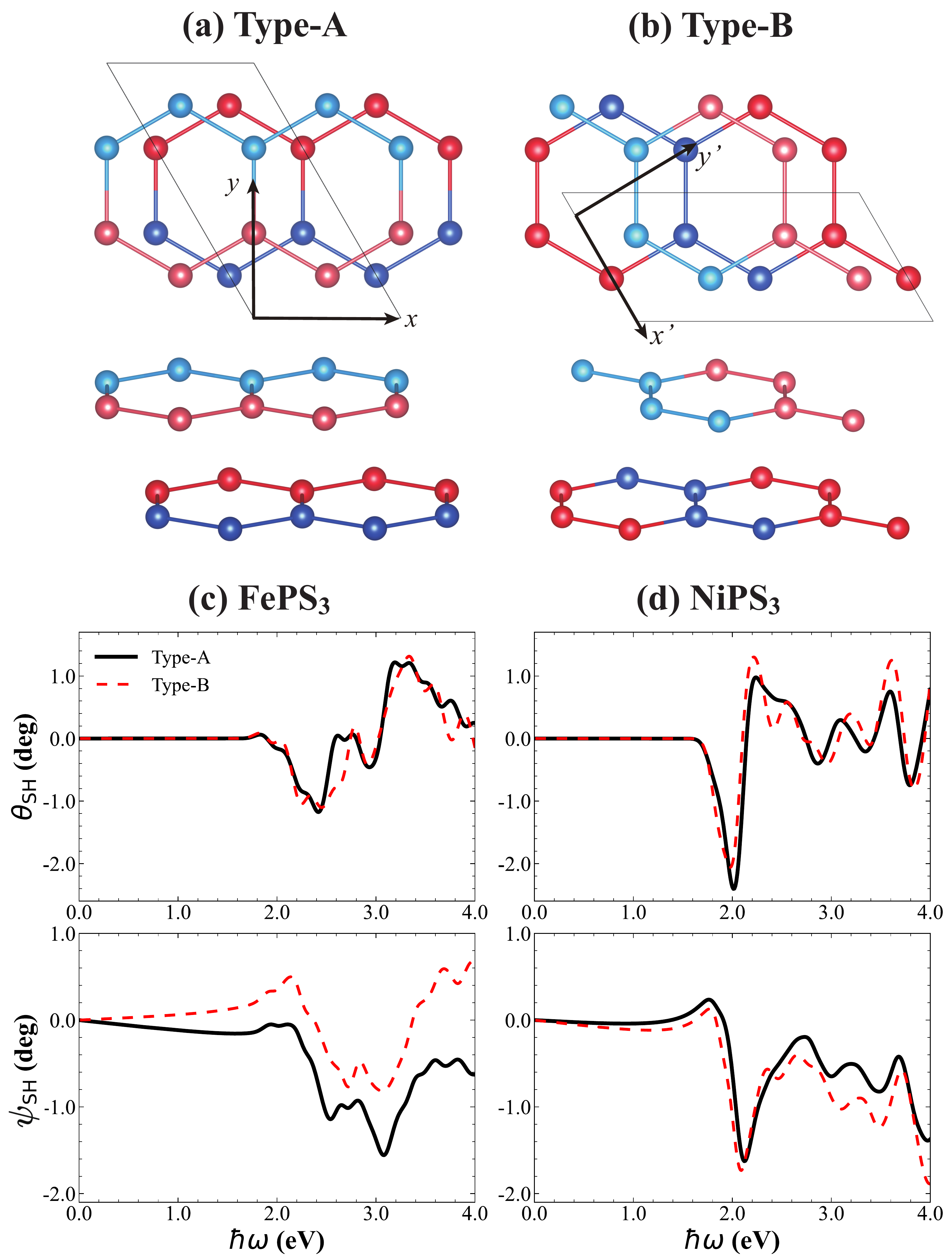}
	\caption{ (a,b) Two types of magnetic structures for bilayer \textit{M}PS$_3$ with the zigzag-AFM chains along the $x$ and $x^{\prime}$-axes.  Bright (dark) red and blue spheres denote the \textit{M} atoms on bottom (top) layer with opposite spin magnetic moments, whereas P and S atoms are not shown.  The solid black lines draw out the 2D primitive cell.  (c,d) Magneto-optical Sch\"{a}fer-Hubert spectra ($\theta_{\rm SH}$ and $\psi_{\rm SH})$ of bilayer FePS$_3$ and NiPS$_3$ with the type-A and type-B magnetic structures. }
	\label{fig:2L}
\end{figure}

Next, we move on to discuss the MOSHE in bilayer FePS$_3$ and NiPS$_3$ on their magnetic ground states.  For FePS$_3$, two types of interlayer magnetic structures have long been reported.  One is the zigzag-AFM chain along the $x$-axis (type-A) with AFM interlayer coupling (Fig.~\ref{fig:2L}a)~\cite{Kurosawa1983}, while the other one is the zigzag-AFM chain along the $x^{\prime}$-axis (type-B) with FM interlayer coupling (Fig.~\ref{fig:2L}b)~\cite{Le-Flem1982}.  Recently, the coexistence of the two types of magnetic structures in multilayer FePS$_3$ has been confirmed by combining MLD and second-harmonic generation measurements~\cite{ZL-Ni2022}.  For NiPS$_3$ powder and single crystal, as far as we know, only the type-A zigzag chain with FM interlayer coupling has been reported~\cite{Brec1986,Wildes2015}.  We speculate that the type-B structure may also exist in bilayer and multilayer NiPS$_3$.

Here we consider both FM and AFM interlayer coupling for type-A and type-B zigzag chains in bilayer FePS$_3$ and NiPS$_3$.  The optical conductivities are not shown because they retain the overall trend in monolayers (Fig.~\ref{fig:SH_1L}b,c) with slightly change in magnitudes due to the weak interlayer vdW interactions.  The calculated SH spectra are plotted in Fig. \ref{fig:2L}c,d, in which the interlayer FM and AFM coupling are not labeled since their spectra are identical to each other.  One can observe that for both FePS$_3$ and NiPS$_3$, the profiles of SH rotation angles for two types of zigzag chains resemble to each other (top panels of Fig.~\ref{fig:2L}c,d).  Moreover, the SH rotation angles for monolayers and bilayers are also similar in the sense that their peaks appear at almost the same photon energy.  The calculated SH rotation angles of bilayer FePS$_3$ (NiPS$_3$) are surprisingly large recording to -1.2$^{\circ}$ at 2.4 eV and 1.2$^{\circ}$ at 3.2 eV (-2.4$^{\circ}$ at 2.0 eV and 1.0$^{\circ}$ at 2.2 eV).  In contrast to the SH rotation angles, the SH ellipticities are highly correlated to the zigzag chain structures (bottom panels of Fig.~\ref{fig:2L}c,d).  The ellipticity spectra of bilayer FePS$_{3}$ with type-A and type-B structures show a striking contrast in a wide range of photon energy, in particular at 2.4 eV and 3.4 eV where $\psi_{\rm SH}$ for type-B structure are zero.  As well, there is significant difference between the type-A and type-B structures of bilayer NiPS$_{3}$ from 3.0 eV to 3.7 eV.  We suggest that the dramatic features in SH ellipticity can be used to distinguish the magnetic structures of bilayer \textit{M}PS$_3$.

In summary, our work establishes a simple theoretical framework for studying the magneto-optical Sch\"{a}fer-Hubert effect in 2D magnetic materials using first-principles calculations, and also proposes second-order magneto-optical spectroscopy to be a powerful technique for accurately detecting the in-plane anisotropy in various magnetic structures.  The calculated results demonstrate that monolayer FePS$_3$ and NiPS$_3$ with the zigzag antiferromagnetic order exhibit large Sch\"{a}fer-Hubert angles (up to 1$^{\circ}$) in visible light and near ultraviolet range.  We further find that the Sch\"{a}fer-Hubert effect is interestingly insensitive to the orientations of N\'{e}el vector.  Finally, the magneto-optical response for bilayer FePS$_3$ and NiPS$_3$ with different stackings of zigzag antiferromagnetic chains are studied.  Surprisingly, the Sch\"{a}fer-Hubert angle of bilayer NiPS$_3$ records up to 2.4$^{\circ}$, and the obvious discrepancy in ellipticity spectra enable a distinction of different interlayer magnetic structures.  The excellent properties render \textit{M}PS$_3$ family a novel AFM materials platform for nanophotonic devices. More importantly, our theoretical framework allows for high-throughput study of Sch\"{a}fer-Hubert effect among 2D AFM materials for finding potentially interesting systems.

	\vspace{0.3cm}
	\noindent{\bf METHODS}
	
	\noindent {\bf First-principles calculations}
	
	\noindent The electronic structure calculations were performed using the projector augmented wave (PAW) method~\cite{Blochl1994}, implemented in Vienna \textit{ab initio} Simulation Package (\textsc{VASP})~\cite{Kresse1999}. The exchange-correlation effects were treated using the generalized gradient approximation with the Perdew-Burke-Ernzerhof (GGA-PBE) parameterization~\cite{Perdew1996}.  The cutoff energy was set to 300 eV and the energy convergence criterion was chosen to 10$^{-6}$ eV.  A $k$-mesh of $12\times12\times1$ ($12\times6\times1$) was used for the $1\times1$ ($1\times2$) unit cell.  Spin-orbit coupling (SOC) effect was included in our calculations.  The correlation effects of the $d$-orbitals of Fe, Ni, and Mn atoms were treated by the GGA+U method~\cite{Dudarev1998}, and the effective Hubbard parameters were set to 3.0 eV, 4.0 eV, and 5.0 eV, respectively.  The experimental lattice constants are adopted for MnPS$_3$ (6.077 \AA), FePS$_3$ (5.947 \AA), and NiPS$_3$ (5.812 \AA)~\cite{Ouvrard1985}.  The van der Waals interactions were considered using the DFT-D2 method~\cite{Grimme2006}.  A vacuum layer of 15 {\AA} was used to eliminate the interactions between the adjacent atomic layers.

	\vspace{0.3cm}
	\noindent{\bf Magneto-optical Sch\"{a}fer-Hubert effect}
	
	\noindent The complex Sch\"{a}fer-Hubert angle in two-dimensional (2D) materials were computed according to Eq.~(\ref{2D-SH-optc}). We constructed the maximally-localized Wannier functions, including the $d$-orbitals of Mn, Fe, and Ni atoms, the $s$- and $p$-orbitals of P atoms, and the $p$-orbitals of S atoms, using \textsc{wannier90} package~\cite{Pizzi2020}.  Then the optical conductivity was calculated using the Kubo-Greenwood formula~\cite{Yates2007},
\begin{eqnarray}\label{eq:OPC}
	\sigma_{\mu\nu}&=& \frac{ie^2\hbar}{N_k V}\sum_{\textbf{k}}\sum_{n, m}\frac{f_{m\textbf{k}}-f_{n\textbf{k}}}{E_{m\textbf{k}}-E_{n\textbf{k}}} \nonumber\\
	&&\times\frac{\langle\psi_{n\textbf{k}}|\hat{\upsilon}_{\mu}|\psi_{m\textbf{k}}\rangle\langle\psi_{m\textbf{k}}|\hat{\upsilon}_{\nu}|\psi_{n\textbf{k}}\rangle}{E_{m\textbf{k}}-E_{n\textbf{k}}-(\hbar\omega+i\eta)},
\end{eqnarray}
where $f_{n\textbf{k}}$, $V$, $N_k$, $\omega$, and $\eta$ are the Fermi-Dirac distribution function, volume of unit cell, total number of $k$-points in the Brillouin zone, photon frequency, and energy smearing parameter, respectively. $\hat{\upsilon}_{\mu(\nu)}$ is velocity operator with subscripts $\mu,\nu \in\{x, y, z\} $ denotes Cartesian components.  $\psi_{n\textbf{k}}$ and $E_{n\textbf{k}}$ are the Wannier functions and interpolated energy at the band index $n$ and momentum $\textbf{k}$, respectively.  A $k$-mesh of $400 \times 400 \times 1$ was used to converge the optical conductivity and $\eta$ was set to be 0.1 eV.  The effective thicknesses ($d$) of MnPS$_{3}$, FePS$_{3}$, and NiPS$_{3}$ monolayers were taken from the interlayer distances of their bulk compounds, that is, 6.796 {\AA}, 6.722 {\AA}, and 6.632 {\AA}, respectively~\cite{Le-Flem1982}.  The experimental refractive index of SiO$_2$ at different photon energies~\cite{Ghosh1999} was acquired from an online database (https://refractiveindex.info).

	\vspace{0.3cm}
	\noindent{\bf Dipole selection rules}
	
	\noindent The characters of the energy bands at high-symmetry $k$-points were determined using MagVasp2trace code~\cite{Y-Xu2020,Elcoro2021}.  The corresponding irreducible corepresentations and dipole selection rules were identified by MSGCorep package~\cite{G-Liu2021,G-Liu2022}.  Here, we take the magnetic space group $P\overline{3}'1m$ as an example to illustrate how to find out the dipole selection rules.  For an in-plane polarized light (i.e., $\textbf{E} \perp z$), the dipole operators are defined by either $-e\hat{x}$ or $-e\hat{y}$, which together transform as the bases of the irreducible corepresentation $\Gamma_{3}$ of the group $P\overline{3}'1m$.  Using the command "showMSGCorepDirectProduct" in MSGCorep package, we can obtain the direct products and their decompositions between $\Gamma_{3}$ and other corepresentations (Supplementary Fig.~3).  It is easy to find that the dipole selection rules are
$\Gamma_4 \leftrightarrow \Gamma_4$ and $\Gamma_4 \leftrightarrow \Gamma_{5}\Gamma_{6}$.

	\vspace{0.3cm}
	\noindent{\bf DATA AVAILABILITY}
	
	\noindent{The data that support the findings of this study are available from the corresponding author on reasonable request.}
	
	\vspace{0.3cm}
	\noindent{\bf CODE AVAILABILITY}
	
	\noindent{The codes that are necessary to reproduce the findings of this study are available from the corresponding author on reasonable request.}

	\vspace{0.3cm}
	\def\bibsection{\noindent{\bf REFERENCES}}


\begin{thebibliography}{10}
\makeatletter
\renewcommand\@biblabel[1]{#1.}
\expandafter\ifx\csname url\endcsname\relax
  \def\url#1{\texttt{#1}}\fi
\expandafter\ifx\csname urlprefix\endcsname\relax\def\urlprefix{URL }\fi
\providecommand{\bibinfo}[2]{#2}
\providecommand{\eprint}[2][]{\url{#2}}

\bibitem{C-Gong2017}
\bibinfo{author}{Gong, C.} \emph{et~al.}
\newblock \bibinfo{title}{Discovery of intrinsic ferromagnetism in
  two-dimensional van der Waals crystals}.
\newblock \emph{\bibinfo{journal}{Nature}} \textbf{\bibinfo{volume}{546}},
  \bibinfo{pages}{265} (\bibinfo{year}{2017}).

\bibitem{B-Huang2017}
\bibinfo{author}{Huang, B.} \emph{et~al.}
\newblock \bibinfo{title}{Layer-dependent ferromagnetism in a van der Waals
  crystal down to the monolayer limit}.
\newblock \emph{\bibinfo{journal}{Nature}} \textbf{\bibinfo{volume}{546}},
  \bibinfo{pages}{270} (\bibinfo{year}{2017}).

\bibitem{B-Huang2018}
\bibinfo{author}{Huang, B.} \emph{et~al.}
\newblock \bibinfo{title}{Electrical control of 2D magnetism in bilayer
  CrI$_3$}.
\newblock \emph{\bibinfo{journal}{Nat. Nanotechnol.}}
  \textbf{\bibinfo{volume}{13}}, \bibinfo{pages}{544}
  (\bibinfo{year}{2018}).

\bibitem{B-Yang2019}
\bibinfo{author}{Yang, B.}, \bibinfo{author}{Zhang, X.}, \bibinfo{author}{Yang,
  H.}, \bibinfo{author}{Han, X.} \& \bibinfo{author}{Yan, Y.}
\newblock \bibinfo{title}{Strain controlling transport properties of
  heterostructure composed of monolayer CrI$_3$}.
\newblock \emph{\bibinfo{journal}{Appl. Phys. Lett.}}
  \textbf{\bibinfo{volume}{114}}, \bibinfo{pages}{192405}
  (\bibinfo{year}{2019}).

\bibitem{B-Huang2020}
\bibinfo{author}{Huang, B.} \emph{et~al.}
\newblock \bibinfo{title}{Emergent phenomena and proximity effects in
  two-dimensional magnets and heterostructures}.
\newblock \emph{\bibinfo{journal}{Nat. Mater.}} \textbf{\bibinfo{volume}{19}},
  \bibinfo{pages}{1276} (\bibinfo{year}{2020}).

\bibitem{Katayama1988}
\bibinfo{author}{Katayama, T.}, \bibinfo{author}{Suzuki, Y.},
  \bibinfo{author}{Awano, H.}, \bibinfo{author}{Nishihara, Y.} \&
  \bibinfo{author}{Koshizuka, N.}
\newblock \bibinfo{title}{Enhancement of the magneto-optical Kerr rotation in
  Fe/Cu bilayered films}.
\newblock \emph{\bibinfo{journal}{Phys. Rev. Lett.}}
  \textbf{\bibinfo{volume}{60}}, \bibinfo{pages}{1426}
  (\bibinfo{year}{1988}).

\bibitem{Suzuki1992}
\bibinfo{author}{Suzuki, Y.}, \bibinfo{author}{Katayama, T.},
  \bibinfo{author}{Yoshida, S.}, \bibinfo{author}{Tanaka, K.} \&
  \bibinfo{author}{Sato, K.}
\newblock \bibinfo{title}{New magneto-optical transition in ultrathin Fe(100)
  films}.
\newblock \emph{\bibinfo{journal}{Phys. Rev. Lett.}}
  \textbf{\bibinfo{volume}{68}}, \bibinfo{pages}{3355}
  (\bibinfo{year}{1992}).

\bibitem{W-Feng2016}
\bibinfo{author}{Feng, W.}, \bibinfo{author}{Guo, G.-Y.} \&
  \bibinfo{author}{Yao, Y.}
\newblock \bibinfo{title}{Tunable magneto-optical effects in hole-doped
  group-IIIA metal-monochalcogenide monolayers}.
\newblock \emph{\bibinfo{journal}{2D Mater.}} \textbf{\bibinfo{volume}{4}},
  \bibinfo{pages}{015017} (\bibinfo{year}{2017}).

\bibitem{X-Zhou2017}
\bibinfo{author}{Zhou, X.}, \bibinfo{author}{Feng, W.}, \bibinfo{author}{Li,
  F.} \& \bibinfo{author}{Yao, Y.}
\newblock \bibinfo{title}{Large magneto-optical effects in hole-doped blue
  phosphorene and gray arsenene}.
\newblock \emph{\bibinfo{journal}{Nanoscale}} \textbf{\bibinfo{volume}{9}},
  \bibinfo{pages}{17405} (\bibinfo{year}{2017}).

\bibitem{Y-Fang2018}
\bibinfo{author}{Fang, Y.}, \bibinfo{author}{Wu, S.}, \bibinfo{author}{Zhu,
  Z.-Z.} \& \bibinfo{author}{Guo, G.-Y.}
\newblock \bibinfo{title}{Large magneto-optical effects and magnetic anisotropy
  energy in two-dimensional
  ${\mathrm{Cr}}_{2}{\mathrm{Ge}}_{2}{\mathrm{Te}}_{6}$}.
\newblock \emph{\bibinfo{journal}{Phys. Rev. B}} \textbf{\bibinfo{volume}{98}},
  \bibinfo{pages}{125416} (\bibinfo{year}{2018}).

\bibitem{Gudelli2019}
\bibinfo{author}{Gudelli, V.~K.} \& \bibinfo{author}{Guo, G.-Y.}
\newblock \bibinfo{title}{Magnetism and magneto-optical effects in bulk and
  few-layer CrI$_3$: a theoretical GGA + U study}.
\newblock \emph{\bibinfo{journal}{New J. Phys.}} \textbf{\bibinfo{volume}{21}},
  \bibinfo{pages}{053012} (\bibinfo{year}{2019}).

\bibitem{Saidl2017}
\bibinfo{author}{Saidl, V.} \emph{et~al.}
\newblock \bibinfo{title}{Optical determination of the N{\'e}el vector in a
  CuMnAs thin-film antiferromagnet}.
\newblock \emph{\bibinfo{journal}{Nat. Photon.}} \textbf{\bibinfo{volume}{11}},
  \bibinfo{pages}{91} (\bibinfo{year}{2017}).

\bibitem{Z-Zheng2018}
\bibinfo{author}{Zheng, Z.} \emph{et~al.}
\newblock \bibinfo{title}{Magneto-optical probe of ultrafast spin dynamics in
  antiferromagnetic CoO thin films}.
\newblock \emph{\bibinfo{journal}{Phys. Rev. B}} \textbf{\bibinfo{volume}{98}},
  \bibinfo{pages}{134409} (\bibinfo{year}{2018}).

\bibitem{Grigorev2021}
\bibinfo{author}{Grigorev, V.} \emph{et~al.}
\newblock \bibinfo{title}{Optical readout of the N{\'e}el vector in the
  metallic antiferromagnet Mn$_2$Au}.
\newblock \emph{\bibinfo{journal}{Phys. Rev. Appl.}}
  \textbf{\bibinfo{volume}{16}}, \bibinfo{pages}{014037}
  (\bibinfo{year}{2021}).
\newblock

\bibitem{Voigt1908}
\bibinfo{author}{Voigt, W.}
\newblock \emph{\bibinfo{title}{Magneto-und Elektrooptik}}
  (\bibinfo{publisher}{Leipzig, B. G. Teubner}, \bibinfo{year}{1908}).

\bibitem{Schafer1990}
\bibinfo{author}{Sch{\"a}fer, R.} \& \bibinfo{author}{Hubert, A.}
\newblock \bibinfo{title}{A new magnetooptic effect related to non-uniform
  magnetization on the surface of a ferromagnet}.
\newblock \emph{\bibinfo{journal}{Phys. Stat. Sol. (a)}}
  \textbf{\bibinfo{volume}{118}}, \bibinfo{pages}{271}
  (\bibinfo{year}{1990}).

\bibitem{K-Kim2019b}
\bibinfo{author}{Kim, K.} \emph{et~al.}
\newblock \bibinfo{title}{Antiferromagnetic ordering in van der Waals 2D
  magnetic material MnPS$_3$ probed by raman spectroscopy}.
\newblock \emph{\bibinfo{journal}{2D Mater.}} \textbf{\bibinfo{volume}{6}},
  \bibinfo{pages}{041001} (\bibinfo{year}{2019}).

\bibitem{Lee2016}
\bibinfo{author}{Lee, J.-U.} \emph{et~al.}
\newblock \bibinfo{title}{Ising-type magnetic ordering in atomically thin
  FePS$_3$}.
\newblock \emph{\bibinfo{journal}{Nano Lett.}} \textbf{\bibinfo{volume}{16}},
  \bibinfo{pages}{7433} (\bibinfo{year}{2016}).

\bibitem{Lancon2018}
\bibinfo{author}{Lan\c{c}on, D.}, \bibinfo{author}{Ewings, R.~A.},
  \bibinfo{author}{Guidi, T.}, \bibinfo{author}{Formisano, F.} \&
  \bibinfo{author}{Wildes, A.~R.}
\newblock \bibinfo{title}{Magnetic exchange parameters and anisotropy of the
  quasi-two-dimensional antiferromagnet ${\mathrm{NiPS}}_{3}$}.
\newblock \emph{\bibinfo{journal}{Phys. Rev. B}} \textbf{\bibinfo{volume}{98}},
  \bibinfo{pages}{134414} (\bibinfo{year}{2018}).

\bibitem{Q-Zhang2021}
\bibinfo{author}{Zhang, Q.} \emph{et~al.}
\newblock \bibinfo{title}{Observation of giant optical linear dichroism in a
  zigzag antiferromagnet FePS$_3$}.
\newblock \emph{\bibinfo{journal}{Nano Lett.}} \textbf{\bibinfo{volume}{21}},
  \bibinfo{pages}{6938} (\bibinfo{year}{2021}).

\bibitem{ZL-Ni2022}
\bibinfo{author}{Ni, Z.}, \bibinfo{author}{Huang, N.},
  \bibinfo{author}{Haglund, A.~V.}, \bibinfo{author}{Mandrus, D.~G.} \&
  \bibinfo{author}{Wu, L.}
\newblock \bibinfo{title}{Observation of giant surface second-harmonic
  generation coupled to nematic orders in the van der Waals antiferromagnet
  FePS$_3$}.
\newblock \emph{\bibinfo{journal}{Nano Lett.}} \textbf{\bibinfo{volume}{22}},
  \bibinfo{pages}{3283} (\bibinfo{year}{2022}).

\bibitem{XX-Zhang2021}
\bibinfo{author}{Zhang, X.-X.} \emph{et~al.}
\newblock \bibinfo{title}{Spin dynamics slowdown near the antiferromagnetic
  critical point in atomically thin FePS$_{3}$}.
\newblock \emph{\bibinfo{journal}{Nano Lett.}} \textbf{\bibinfo{volume}{21}},
  \bibinfo{pages}{5045} (\bibinfo{year}{2021}).

\bibitem{H-Zhang2022}
\bibinfo{author}{Zhang, H.} \emph{et~al.}
\newblock \bibinfo{title}{Cavity-enhanced linear dichroism in a van der Waals
  antiferromagnet}.
\newblock \emph{\bibinfo{journal}{Nat. Photonics}}
  \textbf{\bibinfo{volume}{16}}, \bibinfo{pages}{311}
  (\bibinfo{year}{2022}).

\bibitem{H-Yuan2015}
\bibinfo{author}{Yuan, H.} \emph{et~al.}
\newblock \bibinfo{title}{Polarization-sensitive broadband photodetector using
  a black phosphorus vertical p-n junction}.
\newblock \emph{\bibinfo{journal}{Nat. Nanotechnol.}}
  \textbf{\bibinfo{volume}{10}}, \bibinfo{pages}{707}
  (\bibinfo{year}{2015}).

\bibitem{Biswas2021}
\bibinfo{author}{Biswas, S.}, \bibinfo{author}{Grajower, M.~Y.},
  \bibinfo{author}{Watanabe, K.}, \bibinfo{author}{Taniguchi, T.} \&
  \bibinfo{author}{Atwater, H.~A.}
\newblock \bibinfo{title}{Broadband electro-optic polarization conversion with
  atomically thin black phosphorus}.
\newblock \emph{\bibinfo{journal}{Science}} \textbf{\bibinfo{volume}{374}},
  \bibinfo{pages}{448} (\bibinfo{year}{2021}).

\bibitem{X-Wang2017}
\bibinfo{author}{Wang, X.} \emph{et~al.}
\newblock \bibinfo{title}{Short-wave near-infrared linear dichroism of
  two-dimensional germanium selenide}.
\newblock \emph{\bibinfo{journal}{J. Am. Chem. Soc.}}
  \textbf{\bibinfo{volume}{139}}, \bibinfo{pages}{14976}
  (\bibinfo{year}{2017}).

\bibitem{X-Wang2021}
\bibinfo{author}{Wang, X.} \emph{et~al.}
\newblock \bibinfo{title}{Spin-induced linear polarization of photoluminescence
  in antiferromagnetic van der Waals crystals}.
\newblock \emph{\bibinfo{journal}{Nat. Mater.}} \textbf{\bibinfo{volume}{20}},
  \bibinfo{pages}{964} (\bibinfo{year}{2021}).

\bibitem{Z-Ni2021}
\bibinfo{author}{Ni, Z.} \emph{et~al.}
\newblock \bibinfo{title}{Imaging the N{\'e}el vector switching in the
  monolayer antiferromagnet MnPSe$_3$ with strain-controlled Ising order}.
\newblock \emph{\bibinfo{journal}{Nat. Nanotechnol.}}
  \textbf{\bibinfo{volume}{16}}, \bibinfo{pages}{782}
  (\bibinfo{year}{2021}).

\bibitem{Zvezdin1997}
\bibinfo{author}{Zvezdin, A.~K.} \& \bibinfo{author}{Kotov, V.~A.}
\newblock \emph{\bibinfo{title}{Modern Magnetooptics and Magnetooptical
  Materials}} (\bibinfo{publisher}{Institute of Physics Publishing, Bristol and
  Philadelphia}, \bibinfo{year}{1997}).

\bibitem{Mertins2001}
\bibinfo{author}{Mertins, H.~C.} \emph{et~al.}
\newblock \bibinfo{title}{Observation of the X-ray magneto-optical Voigt
  effect}.
\newblock \emph{\bibinfo{journal}{Phys. Rev. Lett.}}
  \textbf{\bibinfo{volume}{87}}, \bibinfo{pages}{047401}
  (\bibinfo{year}{2001}).
\newblock

\bibitem{Sivadas2015}
\bibinfo{author}{Sivadas, N.}, \bibinfo{author}{Daniels, M.~W.},
  \bibinfo{author}{Swendsen, R.~H.}, \bibinfo{author}{Okamoto, S.} \&
  \bibinfo{author}{Xiao, D.}
\newblock \bibinfo{title}{Magnetic ground state of semiconducting
  transition-metal trichalcogenide monolayers}.
\newblock \emph{\bibinfo{journal}{Phys. Rev. B}} \textbf{\bibinfo{volume}{91}},
  \bibinfo{pages}{235425} (\bibinfo{year}{2015}).

\bibitem{K-Kim2019a}
\bibinfo{author}{Kim, K.} \emph{et~al.}
\newblock \bibinfo{title}{Suppression of magnetic ordering in XXZ-type
  antiferromagnetic monolayer NiPS$_3$}.
\newblock \emph{\bibinfo{journal}{Nat. Commun.}} \textbf{\bibinfo{volume}{10}},
  \bibinfo{pages}{345} (\bibinfo{year}{2019}).

\bibitem{Basnet2022}
\bibinfo{author}{Basnet, R.} \emph{et~al.}
\newblock \bibinfo{title}{Controlling magnetic exchange and anisotropy by
  nonmagnetic ligand substitution in layered $M\mathrm{P}{X}_{3}$
  ($M=\mathrm{Ni}$, Mn; $X=\mathrm{S}$, Se)}.
\newblock \emph{\bibinfo{journal}{Phys. Rev. Res.}}
  \textbf{\bibinfo{volume}{4}}, \bibinfo{pages}{023256} (\bibinfo{year}{2022}).
\newblock

\bibitem{Chittari2016}
\bibinfo{author}{Chittari, B.~L.} \emph{et~al.}
\newblock \bibinfo{title}{Electronic and magnetic properties of single-layer
  $M\mathrm{P}{X}_{3}$ metal phosphorous trichalcogenides}.
\newblock \emph{\bibinfo{journal}{Phys. Rev. B}} \textbf{\bibinfo{volume}{94}},
  \bibinfo{pages}{184428} (\bibinfo{year}{2016}).

\bibitem{isotropy}
\bibinfo{author}{Stokes, H.~T.}, \bibinfo{author}{Hatch, D.~M.} \&
  \bibinfo{author}{Campbell, B.~J.}
\newblock \bibinfo{title}{Isotropy software suite.}
\newblock \bibinfo{journal}{https://stokes.byu.edu/iso/isotropy.php}

\bibitem{Zelezny2017a}
\bibinfo{author}{{\v Z}elezn{\'y}, J.} \emph{et~al.}
\newblock \bibinfo{title}{Spin-orbit torques in locally and globally
  noncentrosymmetric crystals: antiferromagnets and ferromagnets}.
\newblock \emph{\bibinfo{journal}{Phys. Rev. B}} \textbf{\bibinfo{volume}{95}},
  \bibinfo{pages}{014403} (\bibinfo{year}{2017}).

\bibitem{Zelezny2018a}
\bibinfo{title}{A code for analyzing the shape of linear response tensors}.
\newblock \bibinfo{journal}{https://bitbucket.org/zeleznyj/linear-response-symmetry}

\bibitem{S-Kim2018}
\bibinfo{author}{Kim, S.~Y.} \emph{et~al.}
\newblock \bibinfo{title}{Charge-spin correlation in van der Waals
  antiferromagnet ${\mathrm{NiPS}}_{3}$}.
\newblock \emph{\bibinfo{journal}{Phys. Rev. Lett.}}
  \textbf{\bibinfo{volume}{120}}, \bibinfo{pages}{136402}
  (\bibinfo{year}{2018}).
\newblock

\bibitem{W-Feng2015}
\bibinfo{author}{Feng, W.}, \bibinfo{author}{Guo, G.-Y.},
  \bibinfo{author}{Zhou, J.}, \bibinfo{author}{Yao, Y.} \&
  \bibinfo{author}{Niu, Q.}
\newblock \bibinfo{title}{Large magneto-optical kerr effect in noncollinear
  antiferromagnets Mn$_{3}X$ ($X$ = Rh, Ir, Pt)}.
\newblock \emph{\bibinfo{journal}{Phys. Rev. B}} \textbf{\bibinfo{volume}{92}},
  \bibinfo{pages}{144426} (\bibinfo{year}{2015}).
\newblock

\bibitem{X-Yang2021}
\bibinfo{author}{Yang, X.}, \bibinfo{author}{Zhou, X.}, \bibinfo{author}{Feng,
  W.} \& \bibinfo{author}{Yao, Y.}
\newblock \bibinfo{title}{Strong magneto-optical effect and anomalous transport
  in the two-dimensional van der Waals magnets
  ${\mathrm{Fe}}_{n}{\mathrm{GeTe}}_{2}$ ($n=3$, 4, 5)}.
\newblock \emph{\bibinfo{journal}{Phys. Rev. B}}
  \textbf{\bibinfo{volume}{104}}, \bibinfo{pages}{104427}
  (\bibinfo{year}{2021}).

\bibitem{Stroppa2008}
\bibinfo{author}{Stroppa, A.}, \bibinfo{author}{Picozzi, S.},
  \bibinfo{author}{Continenza, A.}, \bibinfo{author}{Kim, M.} \&
  \bibinfo{author}{Freeman, A.~J.}
\newblock \bibinfo{title}{Magneto-optical properties of (Ga,Mn)As: an
  \textit{ab initio} determination}.
\newblock \emph{\bibinfo{journal}{Phys. Rev. B}} \textbf{\bibinfo{volume}{77}},
  \bibinfo{pages}{035208} (\bibinfo{year}{2008}).

\bibitem{Kurosawa1983}
\bibinfo{author}{Kurosawa, K.}, \bibinfo{author}{Saito, S.} \&
  \bibinfo{author}{Yamaguchi, Y.}
\newblock \bibinfo{title}{Neutron diffraction study on MnPS$_3$ and FePS$_3$}.
\newblock \emph{\bibinfo{journal}{J. Phys. Soc. Jpn.}}
  \textbf{\bibinfo{volume}{52}}, \bibinfo{pages}{3919}
  (\bibinfo{year}{1983}).

\bibitem{Le-Flem1982}
\bibinfo{author}{Le~Flem, G.}, \bibinfo{author}{Brec, R.},
  \bibinfo{author}{Ouvard, G.}, \bibinfo{author}{Louisy, A.} \&
  \bibinfo{author}{Segransan, P.}
\newblock \bibinfo{title}{Magnetic interactions in the layer compounds
  $M\mathrm{P}X_3$ ($M$ = Mn, Fe, Ni; $X$ = S, Se)}.
\newblock \emph{\bibinfo{journal}{J. Phys. Chem. Solids}}
  \textbf{\bibinfo{volume}{43}}, \bibinfo{pages}{455}
  (\bibinfo{year}{1982}).
\newblock

\bibitem{Brec1986}
\bibinfo{author}{Brec, R.}
\newblock \bibinfo{title}{Review on structural and chemical properties of
  transition metal phosphorous trisulfides MPS$_3$}.
\newblock \emph{\bibinfo{journal}{Solid State Ionics}}
  \textbf{\bibinfo{volume}{22}}, \bibinfo{pages}{3} (\bibinfo{year}{1986}).
\newblock

\bibitem{Wildes2015}
\bibinfo{author}{Wildes, A.~R.} \emph{et~al.}
\newblock \bibinfo{title}{Magnetic structure of the quasi-two-dimensional
  antiferromagnet NiPS$_3$}.
\newblock \emph{\bibinfo{journal}{Phys. Rev. B}} \textbf{\bibinfo{volume}{92}},
  \bibinfo{pages}{224408} (\bibinfo{year}{2015}).

\bibitem{Blochl1994}
\bibinfo{author}{Bl\"{o}chl, P.~E.}
\newblock \bibinfo{title}{Projector augmented-wave method}.
\newblock \emph{\bibinfo{journal}{Phys. Rev. B}} \textbf{\bibinfo{volume}{50}},
  \bibinfo{pages}{17953} (\bibinfo{year}{1994}).

\bibitem{Kresse1999}
\bibinfo{author}{Kresse, G.} \& \bibinfo{author}{Joubert, D.}
\newblock \bibinfo{title}{From ultrasoft pseudopotentials to the projector
  augmented-wave method}.
\newblock \emph{\bibinfo{journal}{Phys. Rev. B}} \textbf{\bibinfo{volume}{59}},
  \bibinfo{pages}{1758} (\bibinfo{year}{1999}).

\bibitem{Perdew1996}
\bibinfo{author}{Perdew, J.~P.}, \bibinfo{author}{Burke, K.} \&
  \bibinfo{author}{Ernzerhof, M.}
\newblock \bibinfo{title}{Generalized gradient approximation made simple}.
\newblock \emph{\bibinfo{journal}{Phys. Rev. Lett.}}
  \textbf{\bibinfo{volume}{77}}, \bibinfo{pages}{3865}
  (\bibinfo{year}{1996}).

\bibitem{Dudarev1998}
\bibinfo{author}{Dudarev, S.~L.}, \bibinfo{author}{Botton, G.~A.},
  \bibinfo{author}{Savrasov, S.~Y.}, \bibinfo{author}{Humphreys, C.~J.} \&
  \bibinfo{author}{Sutton, A.~P.}
\newblock \bibinfo{title}{Electron-energy-loss spectra and the structural
  stability of nickel oxide: an LSDA+U study}.
\newblock \emph{\bibinfo{journal}{Phys. Rev. B}} \textbf{\bibinfo{volume}{57}},
  \bibinfo{pages}{1505} (\bibinfo{year}{1998}).

\bibitem{Ouvrard1985}
\bibinfo{author}{Ouvrard, G.}, \bibinfo{author}{Brec, R.} \&
  \bibinfo{author}{Rouxel, J.}
\newblock \bibinfo{title}{Structural determination of some MPS$_3$ layered
  phases (M = Mn, Fe, Co, Ni and Cd)}.
\newblock \emph{\bibinfo{journal}{Mater. Res. Bull.}}
  \textbf{\bibinfo{volume}{20}}, \bibinfo{pages}{1181}
  (\bibinfo{year}{1985}).
\newblock

\bibitem{Grimme2006}
\bibinfo{author}{Grimme, S.}
\newblock \bibinfo{title}{Semiempirical GGA-type density functional constructed
  with a long-range dispersion correction}.
\newblock \emph{\bibinfo{journal}{J. Comput. Chem.}}
  \textbf{\bibinfo{volume}{27}}, \bibinfo{pages}{1787}
  (\bibinfo{year}{2006}).

\bibitem{Pizzi2020}
\bibinfo{author}{Pizzi, G.} \emph{et~al.}
\newblock \bibinfo{title}{Wannier90 as a community code: new features and
  applications}.
\newblock \emph{\bibinfo{journal}{J. Phys.: Condens. Matter}}
  \textbf{\bibinfo{volume}{32}}, \bibinfo{pages}{165902}
  (\bibinfo{year}{2020}).
\newblock

\bibitem{Yates2007}
\bibinfo{author}{Yates, J.~R.}, \bibinfo{author}{Wang, X.},
  \bibinfo{author}{Vanderbilt, D.} \& \bibinfo{author}{Souza, I.}
\newblock \bibinfo{title}{Spectral and fermi surface properties from Wannier
  interpolation}.
\newblock \emph{\bibinfo{journal}{Phys. Rev. B}} \textbf{\bibinfo{volume}{75}},
  \bibinfo{pages}{195121} (\bibinfo{year}{2007}).

\bibitem{Ghosh1999}
\bibinfo{author}{Ghosh, G.}
\newblock \bibinfo{title}{Dispersion-equation coefficients for the refractive
  index and birefringence of calcite and quartz crystals}.
\newblock \emph{\bibinfo{journal}{Opt. Commun.}}
  \textbf{\bibinfo{volume}{163}}, \bibinfo{pages}{95}
  (\bibinfo{year}{1999}).
\newblock

\bibitem{Y-Xu2020}
\bibinfo{author}{Xu, Y.} \emph{et~al.}
\newblock \bibinfo{title}{High-throughput calculations of magnetic topological materials}.
\newblock \emph{\bibinfo{journal}{Nature}} \textbf{\bibinfo{volume}{586}},
  \bibinfo{pages}{702} (\bibinfo{year}{2020}).

\bibitem{Elcoro2021}
\bibinfo{author}{Elcoro, L.} \emph{et~al.}
\newblock \bibinfo{title}{Magnetic topological quantum chemistry}.
\newblock \emph{\bibinfo{journal}{Nat. Commun.}} \textbf{\bibinfo{volume}{12}},
  \bibinfo{pages}{5965} (\bibinfo{year}{2021}).

\bibitem{G-Liu2021}
\bibinfo{author}{Liu, G.-B.}, \bibinfo{author}{Chu, M.},
  \bibinfo{author}{Zhang, Z.}, \bibinfo{author}{Yu, Z.-M.} \&
  \bibinfo{author}{Yao, Y.}
\newblock \bibinfo{title}{Spacegroupirep: a package for irreducible
  representations of space group}.
\newblock \emph{\bibinfo{journal}{Comput. Phys. Commun.}}
  \textbf{\bibinfo{volume}{265}}, \bibinfo{pages}{107993}
  (\bibinfo{year}{2021}).

\bibitem{G-Liu2022}
\bibinfo{author}{Liu, G.-B.}, \bibinfo{author}{Zhang, Z.}, \bibinfo{author}{Yu,
  Z.-M.} \& \bibinfo{author}{Yao, Y.}
\newblock \bibinfo{title}{Msgcorep: a package for corepresentations of magnetic space groups}.
\newblock \bibinfo{journal}{Preprint at https://arxiv.org/abs/2211.10740}
(\bibinfo{year}{2022})

\end{thebibliography}

	\vspace{0.3cm}
	\noindent{\bf ACKNOWLEDGEMENTS}
	
	\noindent The authors thank Shifeng Qian and Run-Wu Zhang for helpful discussion.  This work is supported by the National Key R\&D Program of China (Grant Nos. 2022YFA1402600, 2022YFA1403800, and 2020YFA0308800), the National Natural Science Foundation of China (Grant Nos. 12274027, 11874085, 12274028, and 12061131002), the Science \& Technology Innovation Program of Beijing Institute of Technology (Grant No. 2021CX01020), and the Sino-German Mobility Programme (Grant No. M-0142).

	\vspace{0.3cm}
	\noindent{\bf COMPETING INTERESTS}
		
	\noindent{The authors declare no competing interests.}
	
	\vspace{0.3cm}
	\noindent{\bf AUTHOR CONTRIBUTIONS}
	
	\noindent{W.F. and Y.Y. conceived the research. P.Y. derived the formula and performed the first-principles calculations. P.Y. and G.-B.L. carried out the symmetry analysis. P.Y. and W.F. wrote the manuscript with discussion from all authors}

	\vspace{0.3cm}
	\noindent{\bf ADDITIONAL INFORMATION}
	
	\noindent{{\bf Supplementary information} is available for this paper at https://doi.org/xxxxxx.}
	
	\vspace{0.3cm}	
	\noindent{{\bf Correspondence} and requests for materials should be addressed to W.F.}
	
	\vspace{0.3cm}
	\noindent{{\bf Reprints and permission information} is available at http://www.nature.com/reprints}
	
	\vspace{0.3cm}
	\noindent{{\bf Publisher's note} Springer Nature remains neutral with regard to jurisdictional claims in published maps and institutional affiliations.}

\end{document}